\documentclass[%
 reprint,%
 amssymb, amsmath,%
 aip,jap,%
]{revtex4-1}

\usepackage{graphicx}%
\usepackage{bm}%
\usepackage{gensymb}%
\usepackage{amsmath}%
\usepackage{natbib}%
\usepackage[colorlinks=true,linkcolor=blue]{hyperref}%
\expandafter\ifx\csname package@font\endcsname\relax\else
 \expandafter\expandafter
 \expandafter\usepackage
 \expandafter\expandafter
 \expandafter{\csname package@font\endcsname}%
\fi
\bibstyle{aipauth4-1.bst}

\begin{document}

\title{Laser-controlled field effect in graphene/hexagonal boron nitride heterostructures}%

\author{I. Wlasny}%
\email{igor.wlasny@fuw.edu.pl}
\affiliation{Institute of Experimental Physics, Faculty of Physics, University of Warsaw, Pasteura 5, 02-093 Warsaw, Poland}%

\author{R. Stepniewski}%
\affiliation{Institute of Experimental Physics, Faculty of Physics, University of Warsaw, Pasteura 5, 02-093 Warsaw Poland}%

\author{Z. Klusek}%
\affiliation{Department of Solid State Physics, Faculty of Physics and Applied Informatics, University of Lodz, Pomorska 149/153, 90-236 Lodz, Poland}%

\author{W. Strupinski}%
\affiliation{Institute of Electronic Materials Technology, Wolczynska 133, 01-919 Warsaw, Poland}%
\affiliation{Faculty of Physics, Warsaw University of Technology (WUT), Koszykowa 75, 00-662 Warsaw, Poland}%

\author{A. Wysmolek}%
\affiliation{Institute of Experimental Physics, Faculty of Physics, University of Warsaw, Pasteura 5, 02-093 Warsaw, Poland}%

\date{\today}%
\revised{\today}%

\begin{abstract}
The possibility of modification of the local properties of hexagonal boron nitride (h-BN) by laser irradiation is investigated. Investigations conducted using both Raman spectroscopy and electrostatic force microscopy were performed. Laser light induced modifications are found to cause no structural changes. However, they impact the Raman spectra and local charge state of the material. They are also shown to be stable in time and during electrical grounding of the sample. A mechanism of photoionization of deep defects present in h-BN  is proposed to explain the observed phenomenon. The discussed effect opens up a new method of nanostructuration of h-BN based planar heterostructures.
\end{abstract}

\maketitle

\tableofcontents

\section{Introduction}

Two-dimensional van der Waals heterostructures, in recent years, attract growing attention of the scientific and industry communities\cite{Nematian2012,Lu2017,Yeh2017,Wlasny2016}. They owe their popularity to the possibility of creating planar nanodevices, which is related to the characteristics of van der Waals materials as well as the possibility of tailoring their electrical and optical properties\cite{Winther2017,Rogala2016}. This shows their high potential for application in various electronic or optoelectronic devices. In particular, one of the most thoroughly investigated heterostructures, composed of graphene and hexagonal boron nitride, finds use in e.g. tunneling diodes \cite{Lee2014a} or transistors\cite{Petrone2015}.

Graphene is a carbon allotrope with atoms arranged in a hexagonal lattice\cite{Geim2007}. From the point of view of its electronic structure it is a zero-bandgap semiconductor with exceptionally high charge carrier mobility\cite{CastroNeto2009,Buron2015}. However, in relation to the heterostructure construction, the most important aspect of graphene is that its properties are sensitive to a broad spectrum of factors, such as growth method\cite{Shin2011}, interaction with a substrate\cite{Giovannetti2008} or the environmental interactions, such as an electric field effect\cite{Schwierz2010}. 

Hexagonal boron nitride (h-BN) has a similar structure to that of graphene\cite{Hu2011}, including the lattice configuration as well as a closely matched lattice constant. It is, however, a semiconductor with a wide bandgap of over 5\ eV\cite{Jeong2015}. As an ultra-flat dielectric, free of surface charge non-uniformities \cite{Dean2010,Kim2012}, it interacts weakly with graphene layers within a heterostructure. Due to aforementioned slight lattice mismatch and possible rotation periodic superstructures with new functionalities could be formed\cite{Yankowitz2012}.

In this article we investigate the possibility of controlling the 2D materials by means of modification of the h-BN charge state using a focused light beam. In our study we have investigated h-BN samples both exfoliated from bulk crystals as well as transferred CVD-grown (CVD - chemical vapor deposition). Our Raman scattering (RS) and electrostatic force microscopy (EFM) investigations indicate the possibility of locally influencing the properties of the hexagonal boron nitride as well as prove that laser-based experimental techniques should be used with caution when investigating two-dimensional materials. Our studies also allowed us to pinpoint the origin of the observed changes as well as to study their dynamics. Our investigations of graphene/h-BN heterostructure show that the photoionization of deep defect centers has an effect on the charge carrier concentration of graphene due to a field-effect. This is particularly important, as the same effect occurs within heterostructures based on other 2D materials. The discussed phenomenon opens up a new method of nanostructuration of planar heterostructures based on h-BN as well as provides new possibilities of graphene/h-BN structure applications. 

\section{Experimental details}

\subsection{Sample preparation}

The investigations presented in this article were conducted on a series of samples, allowing us to investigate both the effect of laser illumination on the characteristics of standalone h-BN layers and graphene/h-BN heterostructures, which were used to investigate the effect of h-BN modification on graphene. 

The first sample, referred to as h-BN/SiO$_{2}$ throughout the text, was prepared by mechanical exfoliation of commercially available h-BN single crystals\cite{Novoselov2004}. The initial exfoliation was performed using a standard dicing tape (Microworld M07). The final exfoliation and transfer onto a target substrate was performed with 0.1\ mm thick PDMS (polydimethylosiloxane) film. Flakes were deposited onto a boron doped (p-type) silicon substrate (resistivity of 1$\Omega$ cm) with 90\ nm layer of the thermal oxide. Such substrates enhance the optical contrast of the h-BN flakes allowing for their quick identification\cite{Gorbachev2011}. The PDMS film was removed from SiO$_{2}$ during the cooling phase after previously heating the sample up to 80\degree C, which increased the effectiveness of the deposition\cite{Uwanno2015}.

The sample referred to as h-BN/G/SiC served as a model for a heterostructure of graphene and hexagonal boron nitride. It was prepared using the[previously described exfoliation method. However, instead of a SiO$_{2}$/Si substrate, a 4H-SiC crystal with epitaxial graphene monolayer\cite{Strupinski2011} has been used. The graphene layer was grown using a commercial horizontal CVD hot-wall reactor with propane gas as a precursor\cite{Strupinski2011}.

The last sample studied in this article (G/h-BN/SiO$_{2}$) was prepared by the deposition of graphene and hexagonal boron nitride onto the SiO$_{2}$/Si substrate. Both graphene and h-BN were synthesized using the CVD method on high purity Cu foil substrates. They were transferred onto SiO$_{2}$ with a modified wet-transfer method\cite{Suk2011}. Before the etching of copper, a 0.1~mm thick PDMS layer was deposited on the h-BN/Cu. The etching was performed in 0.1~m aqueous solution of iron(III) nitrate nonahydrate at  21\degree C over a time span of 20 hours. PDMS with h-BN/graphene layers was cleaned of the iron(III) nitrate contamination with ten cycles of bathing in deionized water. Subsequently, PDMS with h-BN/graphene was carefully placed on the SiO$_{2}$ substrate, where it was pressed down for 30 minutes. The PDMS layer was removed during the cooling phase after previously heating the sample to 80\degree C.

\subsection{Atomic/Electrostatic Force Microscopy}

Both atomic force microscopy (AFM) and electrostatic force microscopy (EFM) measurements were performed using NT-MDT Ntegra Aura microscope working in atmospheric conditions (air temperature 21\degree C, pressure 1000~hPa), with NT-MDT NSG10/Au cantilevers coated with 35 nm layer of Au. The setup was connected electrically with the sample to provide a positive bias on the cantilever. Such setup allowed both for imaging the topography of the samples in the non-contact mode and EFM measurements. Each of the images was collected with 256 x 256 pts resolution. The results were analyzed using the Gwyddion 2.40 software\cite{David2012}.

\subsection{Raman Spectroscopy and Optical Microscopy}

Sample illumination, Raman measurements and Optical Microscopy were performed using a Renishaw inVia system equipped with Olympus MPLN 100x objective with 100x magnification and and automated XYZ translation stage with 100~nm spatial resolution. A Renishaw RL532C50 single mode laser with a nominal 45 mW output power has been used as an excitation source. Coupled with the optical microscope it allows to create a focused spot with the measured power of 13 mW and diameter of about 500~nm on the surface of the sample. This setup allowed the acquisition of optical images as well as high-resolution Raman measurements and provided the illumination source to the selected areas of the sample. The obtained Raman spectra were analyzed by numerical fitting of the model spectrum based on Lorentzian curves using the Wolfram Mathematica 11.2 software.

\section{Results and discussion}

\subsection{Electric field measurements}

\begin{figure}[h]
\includegraphics{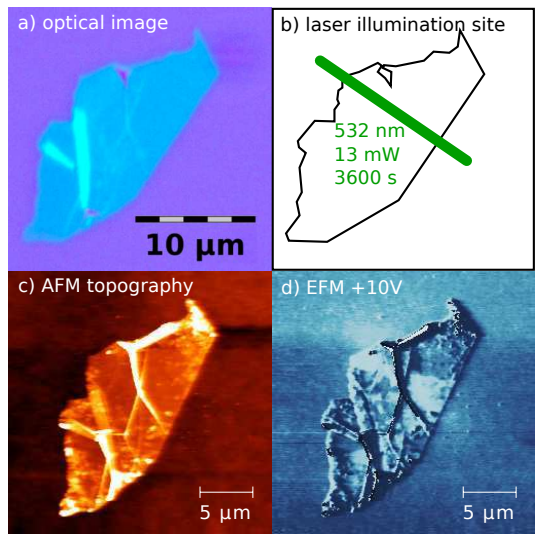}
\centering
\caption{a) Optical image of the h-BN flake on h-BN/SiO$_{2}$ sample, b) Schematic view of the flake with path of the laser illumination indicated with a green line, c) AFM topography of the h-BN flake, d) EFM image of the sample after illumination acquired with 10~V positive bias between AFM cantilever and sample.}
\label{fig-afm}
\end{figure}

The results presented in this section were obtained on the h-BN/SiO$_{2}$ sample. This part of the experiment was performed in order to confirm that the modification is related to the emergence of local electric fields in the sample. To this end a h-BN fragment with large area flat surface was selected. Its optical microscopy image is shown in Figure \ref{fig-afm} a). Basing on the optical contrast, the thickness of the sample is about 5~nm on the terraces with several thicker topographical features \cite{Gorbachev2011}. The thickness and surface characteristics are further confirmed by AFM topography measurements shown in Figure \ref{fig-afm} c).

The sample was illuminated in a linear series of points, as indicated in Figure \ref{fig-afm} b), with conditions that, as we believe, would maximize the observed modification of the local characteristics of the h-BN fragment. Each of the points was separated by 500~nm distance. They were illuminated with 532~nm laser with 13 mW power and a spot of about 500 nm diameter (power density of about 16.5~mW$\mu$m\textsuperscript{-2}) over the time span of 3600 seconds. 

The topography presented in Figure \ref{fig-afm} c) was acquired after the illumination process. It indicates some topographical features that were not seen during the optical measurements - few wrinkles of the h-BN material and slight amount of residues left after the deposition process. However, no distinct features can be found where the illumination was conducted. This clearly shows that despite using the high-power density radiation the sample has not been damaged and shows no visible signs of degradation or deposition of the contamination on the surface. The correlation between the sample illumination and the electric field emergence, however, can be clearly seen in EFM measurements, as presented in Figure \ref{fig-afm} d). In the image a bright area can be seen where the sample was illuminated indicating the presence of electric interaction between the h-BN and the cantilever. This clearly proves that incident focused laser light onto hexagonal boron nitride causes changes in the charge carrier distribution within this material and leads to emergence of local electric fields. As shown later in this article, this effect may be used to control the characteristics of other 2D materials present in the vicinity of the modified h-BN by means of the field-effect with non-invasive method of illumination with high-power density light. Similar effect can also be found after illumination with shorter times and lower power densities (see supplementary material).

\subsection{Time-resolved Raman measurements}

\begin{figure}[h]
\includegraphics{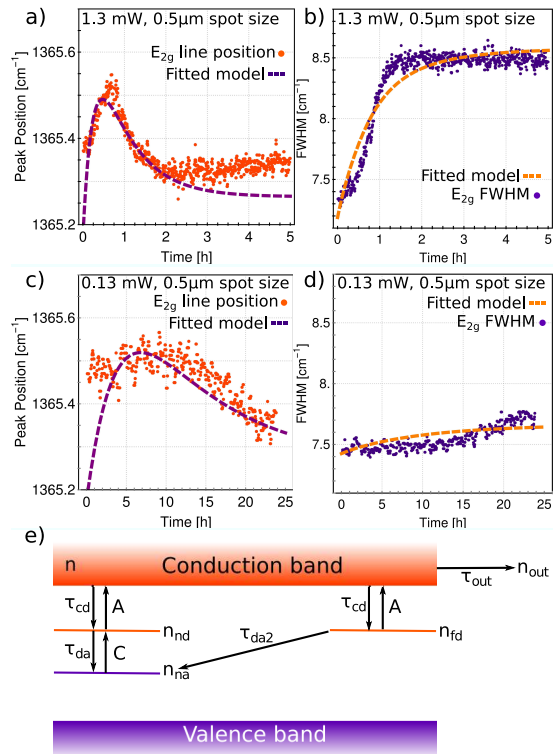}
\centering
\caption{Changes of a, c)~position and b, d)~FWHM (full width at half maximum) of E$_{2g}$ Raman line of h-BN during the illumination with 532~nm laser with of a, b)~1.3~mW power and c, d)~0.13~mW power respectively and 0.5~$\mu$m spot size . Purple, dashed line and orange, dashed line presented in a) and b) present the theoretical model fitted to the results. e) the schematics of the electron transitions between the defect centers in illuminated h-BN assumed in the theoretical model describing the observed photo-ionization of the h-BN.}
\label{fig-pow}
\end{figure}

Results presented in the previous chapter proved that the laser illumination of the hexagonal boron nitride leads to the emergence of local electric fields. However, in order to fully utilize this phenomenon to control other van der Waals materials within the heterostructure it is important to identify the processes behind this phenomenon as well as to investigate the dynamics of the occurring changes. To this end in-situ Raman Spectroscopy measurements were performed on the h-BN/SiO$_{2}$ sample. The exfoliated flake was modified in several points with a focused beam of 532~nm laser. Neutral-density filters were used in order to adjust the power density to the desired levels. The illuminating beam was used to induce the Raman scattering effect for whole duration of the modification process. The time of acquisition for each of the spectra was adjusted to reach a satisfactory signal-to-noise ratio. Each of the spectra has been fitted with a Lorentzian curve, which was used to describe the spectral line related to the E$_{2g}$ vibrational mode of h-BN, which is located at the Raman shift of about 1365~cm\textsuperscript{-1}\cite{Reich2005}. The model results are presented in Figure \ref{fig-pow} a) and b). 

The changes of the position of the E$_{2g}$ line are seen in Figure \ref{fig-pow} a). The initial position of 1365.4~cm\textsuperscript{-1} is related to several factors, such as the thickness or its initial charge state of the h-BN fragment\cite{Gorbachev2011}. During the first hour of the illumination the E$_{2g}$ line is shifted towards higher values, eventually reaching the peak value of about 1365.55~cm\textsuperscript{-1}. After that point, the line is shifted downwards to about 1365.3~cm\textsuperscript{-1} after 2 hours of the process. Subsequently, the line is shifted towards higher values again, albeit at much slower rate. 

The changes can also be seen in the FWHM  (full width at half maximum) of the E$_{2g}$ line (as shown in Figure \ref{fig-pow} b). This parameter starts at about 7.5~cm\textsuperscript{-1} and quickly increases to about 8.5~cm\textsuperscript{-1} after nearly 2 hours into the process. Next, the width of the spectral line seems to be reaching a stable level and does not undergo any changes within the time of observation.

The described behavior can be explained by the modifications of the charge state of deep defects which are induced by the illumination with light. In the proposed model we assume the donors to be of a shallow type and delocalized within the area of several lattice constants and, therefore, with limited coupling to the lattice. On the other hand, strongly localized deep acceptor centers can substantially influence the h-BN structure upon the charge transfer induced by photo-ionization. Thus, we associate the observed shifts of the Raman line energy with the concentration of acceptors in the neutral state\cite{Trautman2003,Kowalski2000,Trautman1992,Leszcznski1991}.

In a simplified model we assume the presence of defect levels\cite{Huang2012} related to acceptors with concentration N$_{A}$ and donors with concentration N$_{D}$+N$_{A}$. We assume that our sample is n-type and in thermodynamic equilibrium concentration of ionized donors and acceptors are the same and equal to N$_{A}$. It was shown\cite{Boguslawski1997} that Coulomb coupling and an additional short-range interaction promote the tendency towards self-compensation by the formation of donor-acceptor pairs. Due to the Coulomb interaction the minimum energy of the system is reached when the donors nearest to the ionized acceptors are ionized. Therefore we divide the donor centers into two groups: "near" - located near acceptor states, initially ionized, with concentration N$_{A}$ and "far" - initially neutral, located far from the acceptors, with concentration of N$_{D}$. The incidence of light onto the crystal may change this equilibrium state. In the simple model, with the excitation below the energy gap, we take into account the excitation that transforms the ionized donor-acceptor pair into a neutral one with probability rate C (Figure \ref{fig-pow} e) and excitations of neutral donors which create free electrons in the conduction band with the probability rate A, the same for both near and far donors (Figure \ref{fig-pow} e)). As recombination processes we take into account donor acceptor recombination for near donors with the recombination time $\tau_{da}$, as well as between far neutral donors and neutral acceptors with significantly bigger recombination time $\tau_{da2}$. Free electrons from the conduction band can, within our model, either be trapped by ionized donor with the recombination time $\tau_{cd}$ or escape outside the illuminated part, with the escape time $\tau_{out}$. We apply a limit N$_{Z}$ to the concentration of electrons that can escape from the illuminated region. To describe this effect we solve a set of rate equations for the described system

\begin{equation}
\label{eq1}
\begin{split}
\frac{dn_{nd}(t)}{dt} = C n_{na} (t) (N_{A}-n_{nd}(t))-\frac{1}{\tau_{da}} n_{nd}(t) (N_{A}- \\
n_{na}(t))-A n_{nd}(t)+\frac{1}{\tau_{cd}}n(t)(N_{A}-n_{nd}(t))
\end{split}
\end{equation}

\begin{equation}
\label{eq2}
\begin{split}
\frac{dn_{fd}(t)}{dt} = -\frac{1}{\tau_{da2}} n_{fd}(t) (N_{A}-n_{na}(t))-A n_{fd}(t)+ \\ 
\frac{1}{\tau_{cd}}n(t)(N_{D}-n_{fd}(t))
\end{split}
\end{equation}

\begin{equation}
\label{eq3}
\begin{split}
\frac{dn_{na}(t)}{dt} = \frac{1}{\tau_{da2}} n_{fd}(t) (N_{A}-n_{na}(t))- \\ 
C n_{na} (t) (N_{A}-n_{nd}(t))+
\frac{1}{\tau_{da}} n_{nd}(t) (N_{A}-n_{na}(t))
\end{split}
\end{equation}

\begin{equation}
\label{eq4}
\begin{split}
\frac{dn(t)}{dt} = A n_{nd}(t)-\frac{1}{\tau_{cd}}n(t)(N_{A}-n_{da}(t))+An_{fd}- \\ 
\frac{1}{\tau_{cd}}n(t)(N_{D}-n_{fd}(t))-\frac{1}{\tau_{out}}n(t)(N_{Z}-n_{out}(t))
\end{split}
\end{equation}

\begin{equation}
\label{eq5}
\begin{split}
\frac{dn_{out}(t)}{dt} = \frac{1}{\tau_{out}}n(t)(N_{Z}-n_{out}(t))
\end{split}
\end{equation}

where:
n$_{nd}$ and n$_{fd}$ are the concentration of electrons present on near and far donors respectively, n is the free electron concentration in the conduction band, n$_{na}$ is the concentration of negatively ionized acceptors, that are treated as the source of electrons that can be excited to a near donor. n$_{out}$ is the concentration of electrons that escape from illuminated part of the sample. 

\begin{table}[ht]
\caption{The initial parameters of the model.}
\centering
\begin{tabular}{| c | c |}
\hline
Parameter & Initial value\\
\hline

n$_{nd}$(0) & 0\\
n$_{fd}$(0) & N$_{D}$\\
n$_{na}$(0) & N$_{A}$\\
n(0) & 0\\
n$_{out}$(0) & 0\\

\hline
\end{tabular}
\label{tab-init}
\end{table}

We assume that each excitation and recombination process is proportional to the concentration of the occupied initial states and to the concentration of empty final states. We do not apply any limits to the concentration of the empty states in the conduction band only. Considering equations \ref{eq1}-\ref{eq5} it is possible to calculate the changes in the occupation in each of the considered states. We solved this set of equations numerically with the initial parameters presented in Tab. \ref{tab-init}.

We assume that the position of the E$_{2g}$ line (E(t)) is tied strictly to the strain of the h-BN lattice. The redistribution of the charge carriers leads to emergence of the non-uniformity of the electric fields within a crystal, which distort the lattice by means of piezoelectricity\cite{Noor-A-Alam2014}. We assume that the changes of the position of the E$_{2g}$ line are related to the number of neutral acceptor centers (N$_{A}$-n$_{na}$(t)), therefore the position of the line can be expressed with equation \ref{eq6}

\begin{equation}
\label{eq6}
E(t) = a(N_{A}-n_{na}(t))+E_{0}
\end{equation}

where E$_{0}$ is the position of the E$_{2g}$ line in the initial state and a is a scaling parameter. It is worth noting that similar correlation between the illumination and the changes in crystalline structure is found in other semiconductors, such as GaN\cite{Trautman2003}, GaAs\cite{Kowalski2000,Trautman1992} or AlGaAs\cite{Leszcznski1991}.

The full width of half maximum of the E$_{2g}$ line is related to the local disorder in the lattice of the illuminated h-BN. We assume the disorder is introduced primarily by the presence of ionized defect centers $(N_{D}-n_{nd}(t))+(N_{D}-n_{fd}(t))+n_{na}(t)$, which create non-uniform electric fields in the crystal leading to rise of the disorder. This effect is lowered by the fact that the electron-dipole interaction depends on the distance between those two objects, thus the dipole contribution can be written as $R\frac{n_{nd}(t)(N_{A}-n_{na}(t))}{N_{A}}$, where R is the parameter corresponding to strength of the dipole contribution into the effect. Therefore the FWHM ($\Sigma$(t)) of the E$_{2g}$ Raman line is given by equation \ref{eq7}

\begin{equation}
\label{eq7}
\begin{split}
\Sigma(t) = c(n_{nd}(t)+n_{fd}(t)+(N_{A}-n_{na}(t))- \\
R\frac{n_{nd}(t)(N_{A}-n_{na}(t))}{N_{A}})+\Sigma_{0}
\end{split}
\end{equation} 

where c is the scaling parameter and $\Sigma_{0}$ is the initial FWHM of the line.

The equations \ref{eq6} and \ref{eq7} were fitted simultaneously to the data presented in figure \ref{fig-pow} a) and b), with the result presented in the same images. Based on this the estimated parameters are presented in Tab. \ref{tab-fit}

\begin{table}[ht]
\caption{The fitting parameters for the model fitted to the changes of Raman line position and FWHM for illumination with 1.3 mW and 0.13 mW 532 nm laser.}
\centering
\begin{tabular}{| c | c | c |}
\hline
Parameter & 1.3~mW, 0.5~$\mu$m spot & 0.13~mW, 0.5~$\mu$m spot\\
\hline

A [h\textsuperscript{-1}] & 0.23 & 0.016\\
C [h\textsuperscript{-1}] & 0.86\textsuperscript{-1} & 0.06\\
$\tau_{cd}$ [h] & 1.5 & 10.5\\
$\tau_{da}$ [h] & 0.62 & 8.7\\
$\tau_{da2}$ [h] & 4.0 & 84\\
$\tau_{out}$ [h] & 2 & 28\\
N$_{A}$ [arb.units] & 1.5 & 1.5\\
N$_{D}$ [arb.units] & 8.5 & 8.5\\
N$_{Z}$ [arb.units] & 0.5 & 2\\

\hline
\end{tabular}
\label{tab-fit}
\end{table}

Our simple model describes the experimental data reasonably well, with the exception to the beginning of the process. This may indicate that the initial state of the system was different than assumed. Furthermore, the number of possible levels associated with the defects in hexagonal boron nitride may be higher than what is taken into account in the model\cite{Huang2012}. Both effects may explain the mentioned discrepancy, particularly if the probabilities of electron transitions are high.

Fitting to the data obtained during the illumination with 532~nm laser with power of 0.13~mW and spot size of 0.5~$\mu$m over a time span of 24 hours, as presented in figure \ref{fig-pow} c) and d) results in the values shown in Table \ref{tab-fit}.

Again, the divergence from the data can be seen for the early stages of the observed process. It seems counterintuitive for the lifetimes, $\tau_{da}$ and $\tau_{da2}$ in particular, to change with the power density of the illuminating light. This fact can be explained, however, by the possibility that the electron transitions may be mediated by the conduction, valence bands or both. In this case the illumination may influence the concentration of the charge carriers which are transported in the sample. Furthermore, changes in illumination may result in a shift of the quasi Fermi level. While this model is simplified and does not take all of the factors into account, such as additional defect center levels, it does explain the major trends in the observed parameters within the investigated scope of the power densities and, therefore, in our opinion, gives an insight on the basic processes behind the changes observed in the experimental results.

While the threshold for the process is not seen in our results, there are more parameters that may have an influence on the process. One of the key aspects is establishing whether interaction between the h-BN layers is influencing the process. This is important both to the analysis of the physical process itself and the applications, where the thickness control issues is vital.

\begin{figure}[h]
\includegraphics{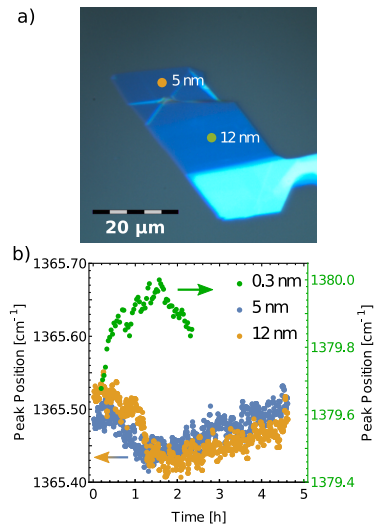}
\centering
\caption{a) Optical microscopy image of the investigated h-BN flake on h-BN/SiO$_{2}$ sample with points points of Raman measurements indicated with green and yellow dot, b) the dynamics of E$_{2g}$ line position changes during the illumination on h-BN of different thicknesses - 0.3~nm (CVD monolayer h-BN, sample G/h-BN/SiO$_{2}$), 5~nm and 12~nm (sample h-BN/SiO$_{2}$).}
\label{fig-thick}
\end{figure}

In order to investigate that aspect we conducted in situ Raman measurements on the exfoliated h-BN flake with areas of different thickness on a h-BN/SiO$_{2}$ sample (see Figure \ref{fig-thick} a) and a monolayer CVD h-BN fragment on G/h-BN/SiO$_{2}$ sample. The changes of the position of the E$_{2g}$ Raman line are presented in Figure \ref{fig-thick} b). It is worth noting that the initial state of the exfoliated flake investigated here, in particular the thicker area, was different from that presented in Figure \ref{fig-pow}, however the changes that can be seen are similar to those previously reported - after 1 hour of illumination the peak position decreases rapidly, and starts increasing slowly after reaching minimum value. The modification of the CVD (0.3~nm thickness) h-BN also seems to be progressing with similar dynamics, however, the last phase is not seen, due to the long acquisition times needed to reach that phase - the local maximum appears after almost 2 hours of the illumination indicating that the photo-excitation occurs at a slower rate in case of this sample. This effect is most likely related to the significant difference in thickness coupled with a low absorption coefficient or the interaction of the boron nitride layers with the substrate. 

\subsection{Modification stability}

The effect of photo-excitation of the charge carriers by the electromagnetic radiation within the hexagonal boron nitride was previously investigated within the scope of the short-term effects, where the changes induced by light were shown not to be stable and disappear after stopping the illumination\cite{Ju2014}. The effect described in this article, however, shows signs of the stability - the electric contrast measurements (seen in Figure \ref{fig-afm} d) were conducted few hours after the illumination, proving that the changes in the charge carrier distribution remain stable at least for that duration. However, in order to establish whether the created modifications will still be seen on the sample after longer times we conducted measurements on the h-BN/SiO$_{2}$ sample 5 months after the illumination. 

\begin{figure}[h]
\includegraphics{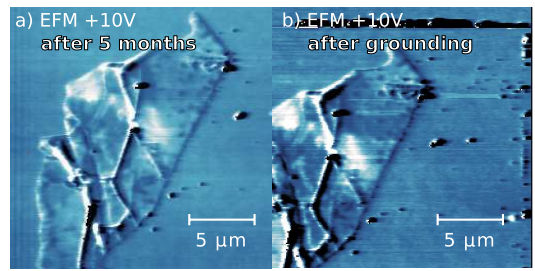}
\centering
\caption{EFM images of the illuminated h-BN flake on h-BN/SiO$_{2}$ sample after a) 5 months of time, b) electrical grounding of the sample.}
\label{fig-stable}
\end{figure}

The EFM image of the sample after that time is shown in Figure \ref{fig-stable} a). It is clearly seen that the contrast has not changed and slight differences between the images in Figure \ref{fig-stable} a) and Figure \ref{fig-pow} d) can be attributed to the differences in the exact state of the used AFM tip and environmental conditions. This shows that the energy levels attributed to the defect centers\cite{Huang2012}, the charge state of which is influenced by the illuminations are deep and may be used to effectively trap the charge for extremely long times. This also proves that the field effect generated by the investigated phenomenon may be used to permanently influence the heterostructure and may indicate the prospective applications of this effect where the stability is vital, such as the non-volatile memories or in nanostructured devices. 

Additionally, attempts to restore the initial state of the illuminated h-BN were performed. This has been done by performing the AFM scans of the surface with the tip with Au coating in the contact mode of AFM. The tip was grounded electrically. After such attempts, however, the EFM contrast can still be seen (see Figure \ref{fig-stable} b). This suggests that the procedure did not have a significant impact on the charge state of the material and observed effects are related to the charge bound to deep defect centers.

\subsection{Response of graphene to h-BN modification}

The last part of the analysis of the phenomenon of light-induced charge carrier photo-excitation in hexagonal boron nitride is the investigation of the influence of the electric fields generated by the effect on graphene within the heterostructures composed of these two materials. The analysis of that effect is basing on Raman Spectroscopy measurements. The spectra were gathered during the illumination conducted on both the heterostructures and the graphene layers on h-BN/G/SiC and G/h-BN/SiO$_{2}$ samples with 1.3 mW 532 nm laser with 0.5 $\mu$m spot size. Lorentzian curves were fitted to the G and 2D bands of the Raman spectra of graphene. Analysis of the parameters of the curves provides information on the charge carrier concentration and the mechanical strain on the material in\cite{Urban2014,Lee2012,Schmidt2011} in case of low n-type doping (under 2~$\cdot $~10\textsuperscript{13}~cm\textsuperscript{-2}) and biaxial strain, which we assume is the case, as no changes in corrugation were observed during AFM measurements (see Fig. \ref{fig-afm} c)) and we see no evidence of a non-uniform character of the observed changes. Within those bounds, the energy of the G band depends on the electron concentration linearly, while 2D is independent from this value. The energies of G and 2D band are related to mechanical strain ($\epsilon$) and electron concentration (n) by equations \ref{eq3b} and \ref{eq4b}\cite{Urban2014}, where $\gamma_{G}$ and $\gamma_{2D}$ are Gr{\"u}neisen parameters for G and 2D bands respectively. $a = $7.38~$\cdot $~10\textsuperscript{13}~cm is a parameter calculated by linear fitting of E$_{G}$(n) function for electrostatic gating and $E_{G}^{0}$ and $E_{2D}^{0}$ are the band positions of the undoped graphene for Raman measurements with 532~nm excitation wavelength. We assumed the $\gamma_{G}/\gamma_{2D}$ to be equal to 0.71\cite{Urban2014}. 

\begin{equation}
\centering
\label{eq3b}
E_{2D} = E_{2D}^{0}-2 \gamma_{2D} \cdot E_{2D}^{0} \cdot \epsilon
\end{equation}

\begin{equation}
\centering
\label{eq4b}
E_{G} = E_{G}^{0}-2 \gamma_{G} \cdot E_{G}^{0} \cdot \epsilon + n \cdot a
\end{equation}

\begin{figure}[h]
\includegraphics{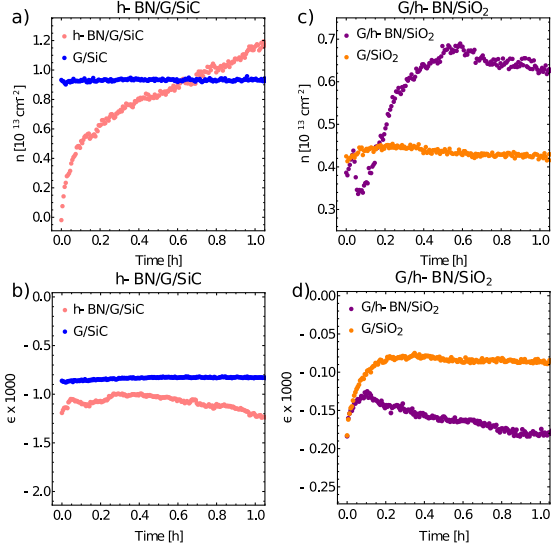}
\centering
\caption{Changes in the a,~c) electron concentration and b,~d) strain in graphene and graphene/h-BN heterostructure in a,~b) h-BN/G/SiC and c,~d) G/h-BN/SiO$_{2}$ samples during illumination with 532 nm 1.3 mW laser with 0.5 $\mu$m. spot size.}
\label{fig-g}
\end{figure}

The results of the measurements are shown in Figure \ref{fig-g}. The differences in the reaction of the graphene and the heterostructures can be clearly seen at first glance. During the entirety of the illumination the strain of the graphene outside the heterostructures, both in case of the epitaxial material on the native substrate (h-BN/G/SiC) and after the transfer (G/h-BN/SiO$_{2}$) show nearly no reaction, as shown in Figure \ref{fig-g} b) and d). The only observed changes are slight and occur within the first 0.2~hours in case of the transferred material. This is most likely associated with the reaction of the non-relaxed interface to the changes in the charge state of hexagonal boron nitride. This also has an impact on the charge carrier concentration in the graphene fragment on the surface. Increase of this parameter may allow the interface between graphene and surface to relax, which is consistent with the results seen in Figure \ref{fig-g} d). The change is very slight, however, reaching only 0.01~\%. 

The major differences appear, however, in the electron doping of the investigated samples (see Figure \ref{fig-g} a) and c)). As was the case for the strain, the graphene outside of the heterostructure shows nearly no reaction to the laser illumination in both of the investigated samples. The exact value of the doping is related to the fact that the material was grown on different substrates leading to the different initial value of charge carrier concentration. In case of the h-BN/G/SiC sample the concentration in graphene was at the level of about 0.9~10\textsuperscript{13}~cm\textsuperscript{-2} (Figure \ref{fig-g} a)), while within the heterostructure this value was changing from 0 to 1.2~10\textsuperscript{13}~cm\textsuperscript{-2} with decreasing rate. It is worth pointing out that the initial doping of the heterostructure is significantly different from standalone graphene (graphene outside the heterostructure) most likely due to the possibility of transfer of static charges to the system during exfoliation and transfer of the h-BN layers. 

The evolution of the charge carrier concentration in G/h-BN/SiO$_{2}$ progresses similarly, however, there are few key differences, as shown in Figure \ref{fig-g} c). First, the change in the electron doping of the heterostructure is much smaller and is contained within the range of 0.35~10\textsuperscript{13}~cm\textsuperscript{-2} to 0.70~10\textsuperscript{13}~cm\textsuperscript{-2}. This fact is understandable considering that the h-BN layer is much thinner and therefore the density of the defect centers is lower. The generated field, is weaker and thus has lesser impact on graphene. The second difference lies in the fact that the electron concentration is decreasing after 0.6~hours. As before that time the character of the changes is similar to that in Figure \ref{fig-g} a). It is likely that the rate of changes is different. Again, this can be attributed to the fact that in the G/h-BN/SiO$_{2}$ sample we are dealing with the monolayer material, which is related to the lower maximum charge carrier concentration in the defect levels.

Basing on the above results, it can be clearly seen that the presence of the h-BN causes the modification of the charge carrier concentration in graphene during the illumination with high power density light. As the reaction of the standalone graphene is significantly lower the changes can be attributed to the emergence of the non-uniform electric fields in h-BN. The above results show that the electron concentration can be changed in wide range (from 0 to 10\textsuperscript{13}~cm\textsuperscript{-2}), therefore it shows that it is possible to control the electric conductivity of the heterostructure using the illumination with high power density light. This opens up new possibilities of application of the effect in data storage or electronic component and circuit engineering. Furthermore, as the photo-excitation effect occurs in h-BN, it may also be used it in tailoring the properties of other 2D materials.

\section{Conclusions}

In summary, it was shown that the illumination of the hexagonal boron nitride leads to the emergence of local electric fields. This effect is attributed to transitions of the electric charge carriers between the levels associated with the structural defects. Those defect levels were proven to be located deep within the band gap of h-BN. Because of this fact the light-induced modification of the charge carrier concentration was shown to be stable within long periods of time. 

The results presented in this article show that the electric fields appearing in h-BN may be used to control the properties of other 2D materials, which can be use to construct the heterostructures with hexagonal boron nitride. In particular, the presented results show the influence of illumination on graphene/h-BN heterostructure, where the electric field may be used to control the electron concentration in graphene.

The described effect allows for easy and non-invasive tailoring of the electronic properties of two-dimensional materials and may open new directions of research of the van der Waals heterostructures and their applications.

\section{Supplementary Material}

See supplementary material for more records of illumination of both h-BN and graphene/h-BN heterostructures as well as selected Raman spectra of the data presented in this article.

\section{Acknowledgments}

This work was supported by National Science Centre project granted on the basis of the decision number DEC-2015/16/S/ST3/00451. The work was also financially supported by National Science Centre project
2015/19/B/ST3/03142 as well as the European Union Seventh Framework Program (Grant. No. 604391 Graphene Flagship).


\section{References}

\begin{thebibliography}{37}

\bibitem{Nematian2012}
H. Nematian, M. Moradinasab, M. Pourfath, M. Fathipour,
and H. Kosina.
\newblock J. Appl. Phys. 111, 093512 (2012).

\bibitem{Lu2017}
S. Lu and A. J. H. McGaughey.
\newblock J. Appl. Phys. 121, 115103 (2017).

\bibitem{Yeh2017}
C.-H. Yeh, Z.-Y. Liang, Y.-C. Lin, T.-L. Wu, T. Fan, Y.-C. Chu, C.-H. Ma, Y.-C. Liu, Y.-H. Chu, K. Suenaga, and P.-W. Chi.
\newblock ACS Appl. Mat. Interfaces 9, 36181 (2017).

\bibitem{Wlasny2016}
I. Wlasny, M. Rogala, P. Dabrowski, P. J. Kowalczyk, A. Busiakiewicz, W. Kozlowski, L. Lipinska, J. Jagiello, M. Aksienionek, Z. Sieradzki, I. Krucinska, M. Puchalski, E. Skrzetuska, Z. Draczynski, and Z. Klusek.
\newblock Mat. Chem. Phys. 181, 409 (2016).

\bibitem{Winther2017}
K. T. Winther and K. S. Thygesen.
\newblock 2D Mat. 2D Materials 4, 025059 (2017).

\bibitem{Rogala2016}
M. Rogala, P. Dabrowski, P. Kowalczyk, I. Wlasny, W. Kozlowski, A. Busiakiewicz, I. Karaduman, L. Lipinska, J. Baranowski, and Z. Klusek.
\newblock Carbon 103, 235-241 (2016).

\bibitem{Lee2014a}
S. H. Lee, M. S. Choi, J. Lee, C. Ho Ra, X. Liu, E. Hwang, J. H. Choi, J. Zhong, W. Chen, and W. J. Yoo
\newblock Appl. Phys. Lett. 104, 053103 (2014).

\bibitem{Petrone2015}
N. Petrone, T. Chari, I. Meric, L. Wang, K. L. Shepard, and J. Hone
\newblock ACS Nano 9 8953 (2015).

\bibitem{Geim2007}
A. K. Geim and K. S. Novoselov.
\newblock Nat. Mater. 6, 183 (2007).

\bibitem{CastroNeto2009}
A. H. Castro Neto, N. M. R. Peres, K. S. Novoselov, and A. K. Geim.
\newblock Rev. Mod. Phys. 81, 109 (2009).

\bibitem{Buron2015}
J. D. Buron, F. Pizzocchero, P. U. Jepsen, D. H. Petersen, J. M. Caridad, B. S. Jessen, T. J. Booth, and P. Bggild.
\newblock Sci. Rep. 5, 12305 (2015).

\bibitem{Shin2011}
Y. J. Shin, R. Stromberg, R. Nay, H. Huang, A. T. Wee, H. Yang, and C. S. Bhatia.
\newblock Carbon 49, 4070 (2011).

\bibitem{Giovannetti2008}
G. Giovannetti, P. A. Khomyakov, G. Brocks, V. M. Karpan, J. Van Den Brink, and P. J. Kelly.
\newblock Phys. Rev. Lett. 101, 026803 (2008).

\bibitem{Schwierz2010}
F~Schwierz.
\newblock Nat. Nanotech. 5, 487 (2010).

\bibitem{Hu2011}
M. L. Hu, J. L. Yin, C. X. Zhang, Z. Yu, and L. Z. Sun.
\newblock J. Appl. Phys. 109, 073708 (2011).

\bibitem{Jeong2015}
H. Jeong, S. Bang, H. M. Oh, H. J. Jeong, S. J. An, G. H. Han, H. Kim, K. K. Kim, J. C. Park, Y. H. Lee, G. Lerondel, and M. S. Jeong.
\newblock ACS Nano 9, 10032 (2015).

\bibitem{Dean2010}
C. R. Dean, A. F. Young, I. Meric, C. Lee, L. Wang, S. Sorgenfrei,
K. Watanabe, T. Taniguchi, P. Kim, K. L. Shepard, and J. Hone.
\newblock Nat. Nanotech. 5, 722 (2010).

\bibitem{Kim2012}
K. K. Kim, A. Hsu, X. Jia, S. M. Kim, Y. Shi, M. Dresselhaus, T. Palacios, and J. Kong.
\newblock Nano Lett. 12, 161–166 (2012).

\bibitem{Yankowitz2012}
M. Yankowitz, J. Xue, D. Cormode, J. D. Sanchez-Yamagishi, K. Watanabe, T. Taniguchi, P. Jarillo-Herrero, P. Jacquod, and B. J. LeRoy.
\newblock Nat. Phys. 8, 382–386 (2012).
  
\bibitem{Novoselov2004}
K. S. Novoselov, A. K. Geim, S. V. Morozov, D. Jiang, Y. Zhang, S. V. Dubonos, I. V. Grigorieva, and A. A. Firsov.
\newblock Science,306, 666 (2004).

\bibitem{Gorbachev2011}
R. V. Gorbachev, I. Riaz, R. R. Nair, R. Jalil, L. Britnell, B. D. Belle, E. W. Hill, K. S. Novoselov, K. Watanabe, T. Taniguchi,
A. K. Geim, and P. Blake.
\newblock Small 7, 465 (2011).

\bibitem{Uwanno2015}
T. Uwanno, Y. Hattori, T. Taniguchi, K. Watanabe, and K. Nagashio.
\newblock 2D Materials 2, 041002 (2015).

\bibitem{Strupinski2011}
W. Strupinski, K. Grodecki, A. Wysmolek, R. Stepniewski, T. Szkopek, P. E. Gaskell, A. Grüneis, D. Haberer, R. Bozek, J. Krupka, and J. M. Baranowski.
\newblock Nano Lett. 11, 1786 (2011).

\bibitem{Suk2011}
J. W. Suk, A. Kitt, C. W. Magnuson, Y. Hao, S. Ahmed, J. An, A. K. Swan, B. B. Goldberg, and R. S. Ruoff.
\newblock ACS Nano 5, 6916 (2011).

\bibitem{David2012}
N. David and K. Petr.
\newblock Open Phys. 10, 181-188 (2012).

\bibitem{Reich2005}
S. Reich, a. Ferrari, R. Arenal, a. Loiseau, I. Bello, and J. Robertson,
\newblock Phys. Rev. B 71, 1 (2005).

\bibitem{Trautman2003}
P. Trautman, K. Pakula, R. Bozek, and J. M. Baranowski,
\newblock Appl. Phys. Lett. 83, 3510 (2003).

\bibitem{Kowalski2000}
G. Kowalski, S. P. Collins, and M. Moore
\newblock J. App. Phys. 87, 3663 (2000).

\bibitem{Trautman1992}
P. Trautman and J. M. Baranowski, 
\newblock Phys. Rev. Lett. 69, 664 (1992).

\bibitem{Leszcznski1991}
M. Leszcznski, T. Suski, and G. Kowalski, 
\newblock Semicond. Sci. Technol. 6, 59 (1991).

\bibitem{Huang2012}
B. Huang and H. Lee, 
\newblock Phys. Rev. B 86, 245406 (2012).

\bibitem{Boguslawski1997}
P. Boguslawski and J. Bernholc, 
\newblock Phys. Rev. B 56, 9496 (1997).

\bibitem{Noor-A-Alam2014}
M. Noor-A-Alam, H. J. Kim, and Y.-H. Shin, 
\newblock Phys. Chem. Chem. Phys. 16, 6575 (2014).
  
\bibitem{Ju2014}
L. Ju, J. Velasco, E. Huang, S. Kahn, C. Nosiglia, H.-Z. Tsai, W. Yang, T. Taniguchi, K. Watanabe, Y. Zhang, G. Zhang, M. Crommie, A. Zettl, and F. Wang.
\newblock Nat. Nanotech. 9, 348 (2014).

\bibitem{Urban2014}
J. M. Urban, P. Dbrowski, J. Binder, M. Kopciuszyski, A. Wysmoek, Z. Klusek, M. Jaochowski, W. Strupiski, and J. M. Baranowski
\newblock J. Appl. Phys. 115, 233504 (2014).

\bibitem{Lee2012}
J. E. Lee, G. Ahn, J. Shim, Y. S. Lee, and S. Ryu, 
\newblock Nat. Comm. 3, 1024 (2012).

\bibitem{Schmidt2011}
D. A. Schmidt, T. Ohta, and T. E. Beechem, 
\newblock Phys. Rev. B 84, 235422 (2011).

\end{thebibliography}
\end{document}